\documentclass[preprint,12pt]{elsarticle}

\usepackage{lineno, hyperref, amsmath}
\modulolinenumbers[5]

\journal{Nuclear Instruments and Methods in Physics Research A}

\begin{document}

\begin{frontmatter}

\title{The e\_LiBANS facility: a new compact thermal neutron source based on a medical electron LINAC}

\author[unito,infn]{V. Monti\corref{corrauth}}
\cortext[corrauth]{Corresponding author}
\ead{valeria.monti@unito.it}

\author[unito,infn]{M. Costa}
\author[unito,infn]{E. Durisi}
\author[unito,infn]{E. Mafucci}
\author[unito,infn]{L. Menzio}
\author[unito,infn]{O. Sans-Planell}
\author[unito,infn]{L. Visca}
\author[lnf]{R. Bedogni}
\author[lnf,uab]{M. Treccani}
\author[polimi]{A. Pola}
\author[polimi]{D. Bortot}
\author[units,infnts]{K. Alikaniotis}
\author[units,infnts]{G. Giannini}
\author[ciemat]{J. M. Gomez-Ros}

\address[unito]{Universit\`a degli Studi di Torino, via P. Giuria 1, 10125, Torino, Italy}
\address[infn]{INFN, Sezione di Torino, via P. Giuria 1, 10125, Torino, Italy}
\address[lnf]{INFN, Laboratori Nazionali di Frascati, via Enrico Fermi 40, 00044, Frascati, Italy}
\address[units]{Universit\'a degli Studi di Trieste, via Valerio 2, 34127 Trieste, Italy }
\address[infnts]{INFN, Sezione di Trieste, via Valerio 2, 34127 Trieste, Italy}
\address[uab]{UAB, Departament de Fisica, Universitat Autónoma de Barcelona, 08193 Bellaterra, Spain}
\address[polimi]{Politecnico di Milano, Dipartimento di Energia, via La Masa 34, Milano, 20156 Italy}
\address[ciemat]{CIEMAT, Av. Complutense 40,28040, Madrid, Spain}

\begin{abstract}
A new photonuclear thermal neutron facility has been developed at the Physics Department of University of Torino. 
The facility is based on a medical electron LINAC coupled to a compact converter and moderator assembly. 
A homogenous thermal neutron field of the order of 10$^6$ cm$^{-2}$s$^{-1}$ is achievable in the  enclosed irradiation cavity with low gamma and fast neutron contaminations. 
Its intensity can be tuned varying the LINAC current.
These characteristics make the source appropriate for several applications like detectors development, material studies and BNCT preclinical research. This work describes the project and the experimental characterization of the facility. This includes the measurement of the thermal neutron fluence rate, the determination of the neutron energy spectrum and of the thermal neutron field uniformity and the evaluation of the gamma contamination. 
\end{abstract}


\end{frontmatter}


\section{Introduction}
The worldwide effort in developing compact and reduced-cost slow neutron sources not based on nuclear fission has raised the interest in accelerator driven neutron sources. These are often based on proton or deuteron beams impinging on light materials \cite{carpenter} with relevant technological and cost-effective investments.
An interesting alternative for neutron production is the use of photonuclear ($\gamma$, n) reaction.
Starting from an electron LINAC, intense thermal neutron beams can be obtained by introducing a photonuclear converter and a moderating assembly \cite{phones}.

An example of such a development is the INFN e\_LiBANS (Electron-LINAC Based Actively monitored Neutron Source) project (2016-2018), within which a thermal neutron source for interdisciplinary applications was established, starting from a 18 MV medical-type electron LINAC. 
The possible applications of this facility are development of new neutron detectors, material studies or preclinical research in Boron Neutron Capture Therapy (BNCT).

This article presents the entire development of the e\_LiBANS thermal neutron source: the design, the construction and the experimental characterization.

\subsection{The LINAC}
\label{sec:linac}

An Elekta SL-18 MV Precise SW electron LINAC, formerly used in a hospital for radiation therapy, was installed at the Physics Department of University of Torino for research purposes.
The accelerator is able to deliver electron or X-ray beams with primary electron energy ranging between 4 and 18 MeV. 
The X-Ray output is the result of a conversion of the primary electron beam on an internal target. The emitted X-rays have a typical bremsstrahlung spectrum with endpoint set to the primary electron energy.  
For the present work, X-ray output at the maximum linac energy has been selected.
In the LINAC head a system of jaws and multileaf collimators is present. For the purposes of this work they are set to produce a (40x40) cm$^2$ photon field at 100 cm from the target along the beam axis.

\subsection{Photonuclear interaction}
\label{sec:ph-n-int}
The neutron production arises from the interaction of the high energetic photons with the nuclei of the target material via ($\gamma$, n) reaction. 
Some heavy metals, like Pb and W, show the highest ($\gamma$, n) cross section values which are around 600 mb, while photonuclear processes are often negligible in light materials. The photon energy threshold is related to the binding energy of the neutron to the nucleus and, indeed, high Z elements show energy thresholds around 6-8 MeV. Emitted neutrons have energy around 1 MeV characterized by a typical evaporative spectrum.

\section{The photoconverter assembly}
\label{sec:project}
\begin{figure}
\centering
  \includegraphics[width=0.65\linewidth]{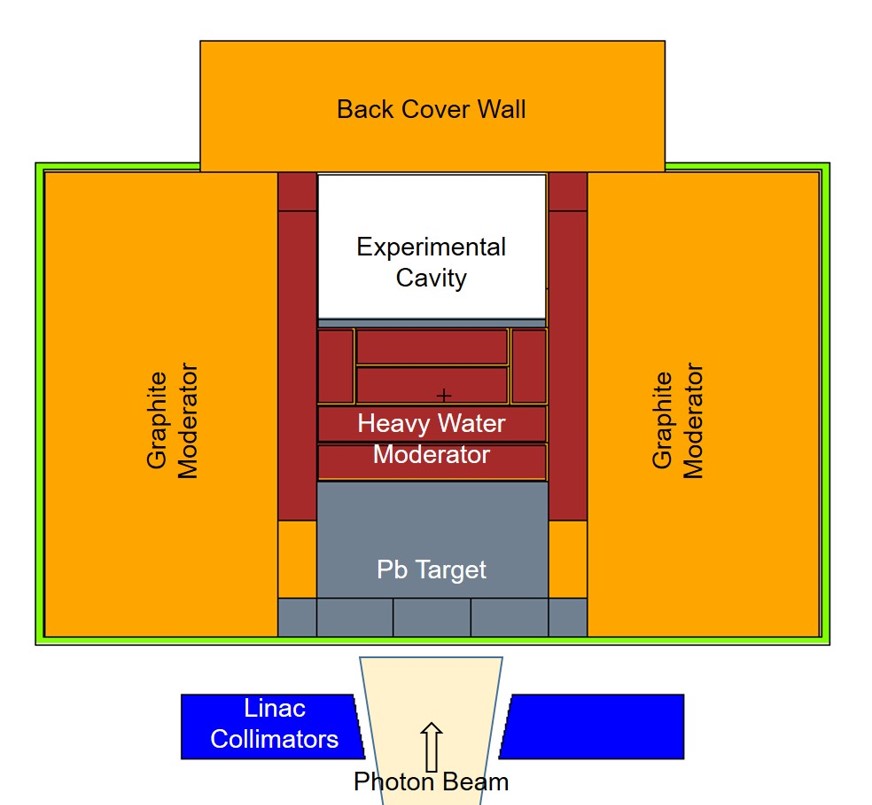}  
 \caption{Longitudinal view of the photoconverter geometry. The X-ray beam, coming from the bottom of the image, impinges on the lead target. Neutrons go through the heavy water and graphite moderator and reach the experimental cavity. A thin slab of polyethylene surrounds the assembly.}
 \label{fig:geom}
\end{figure}

The structure to exploit the photonuclear processes for the thermal neutron production consists of a heavy material conversion target, to produce the neutrons, and of a moderator volume, to slow down the neutrons to thermal energy, embedded in large reflector blocks.
The geometry, which is described in the following paragraphs, has been optimized in order to maximize the thermal neutron production and to minimize the fast neutron and gamma contamination. The gamma background is due to primary unconverted photons and to photons coming from neutron capture processes in the moderator. 
The materials have been chosen after an accurate study of the cross sections, taking into account practical constraints like mechanical properties,
material availability and radiation protection issues. 
The final configuration of the photoconverter, shown in Fig. \ref{fig:geom}, is composed of the following parts.
\begin{itemize}
 \item A thick lead block serves as conversion target and unconverted X-ray absorber. The target transverse section is 40x40 cm$^2$ in the first 5 cm of thickness. The X-Ray beam is intercepted at almost 60 cm from the LINAC conversion target, where it has a transverse section of 25x25 cm$^2$. The photoconverter target transverse section is reduced to 30x30 cm$^2$ in the remaining 15 cm of thickness. The total block thickness is indeed 20 cm, obtained with 4 dovetail joint walls, 5 cm thick. The neutron production takes place mainly in the first 4 cm because upon further penetration
 in the target material the bremsstrahlung photon beam is degraded in energy and intensity and the probability of photonuclear interaction becomes negligible. The extra-thickness of the target is used to absorb the photon beam.
 
 \item The moderator is composed of a central core in heavy water contained in carbon boxes and an external structure made of 30 cm of graphite on each side. A cavity is obtained in the middle of graphite where the samples are exposed for the irradiation. A graphite back cover wall, 30 cm thick, closes the cavity enhancing the field homogeneity through the back-scattering. The cavity section is 30x30 cm$^2$ with 20 cm of longitudinal length.
 
 \item A 0.4 cm thick borated rubber shield, enclosed in thin polyethylene slabs, surrounds the all structure to avoid thermal neutrons diffusion in the experimental room. 
\end{itemize}

 The project of the photoconverter has been carried out through an extensive Monte Carlo simulation study.
 The tool used for the simulation is MCNP6 \cite{mcnp6} which allows the transport of neutrons, electrons and photons in various materials using cross section tables from evaluated data libraries. For this work, cross section data for the atomic interactions are taken from ENDF-B/VII.1 evaluated nuclear data libraries \cite{endf}, while for photonuclear interaction the data are provided by the Los Alamos National Laboratory Nuclear Physics Group. Molecular excitation levels are taken into account for the thermal neutron treatment throughout the special S($\alpha$,$\beta$) cross section tables for graphite, polyethylene and heavy water \cite{thermal-treatment}.

\section{The thermal neutron source}
\label{sec:development}
\begin{figure*}
 \centering
 \includegraphics[width=\linewidth]{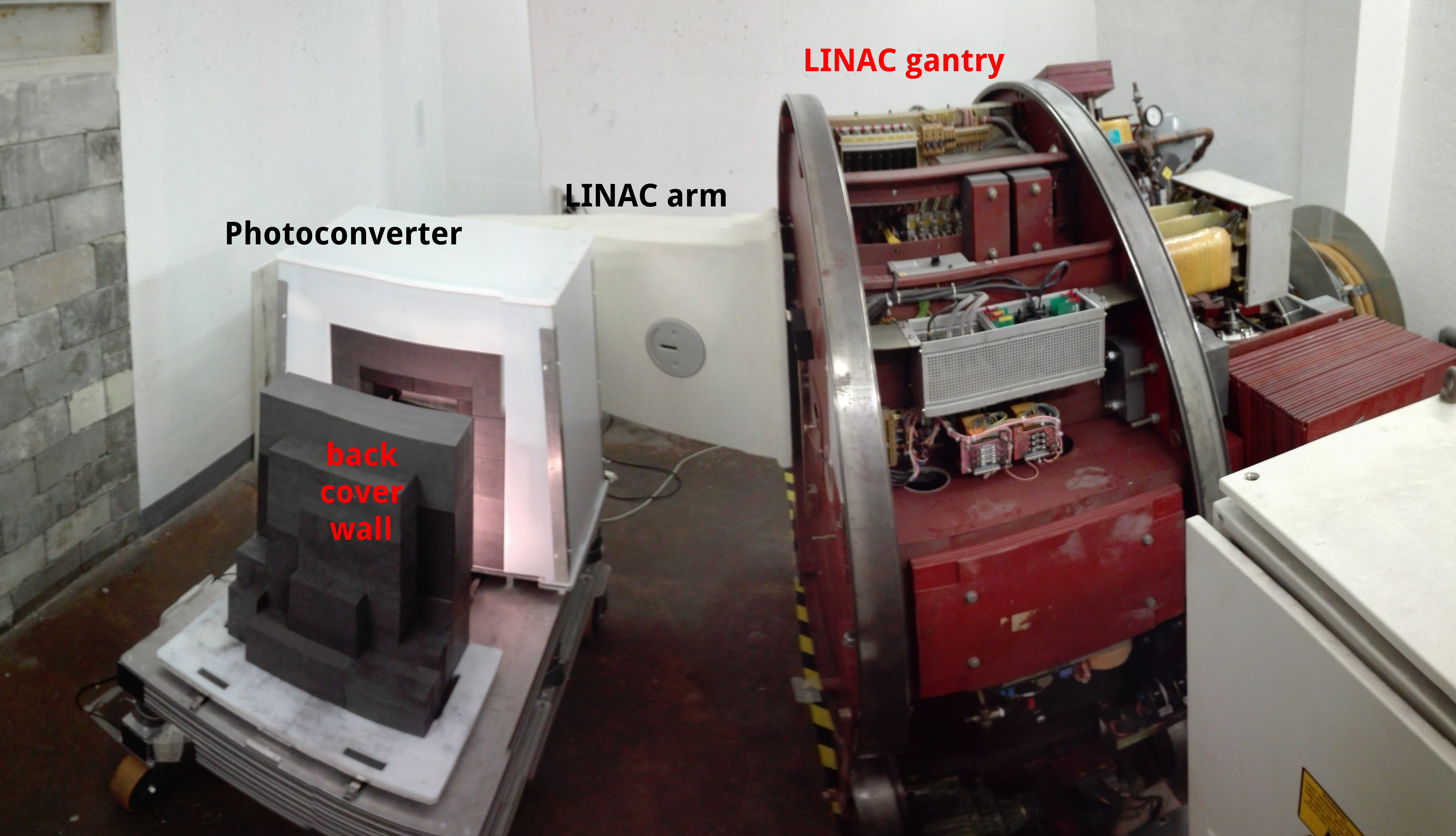}
 \caption{Picture of the e\_LiBANS thermal neutron facility. The main parts, from the left to the right, are: the back cover wall, the photoconverter assembly, the LINAC arm, the LINAC gantry.  The LINAC head is hidden behind the photoconverter and the experimental cavity is hidden behind the back cover wall.} 
  \label{fig:foto-sorgente}
 \end{figure*}

The photoconverter is mounted on a movable support and positioned in front of the LINAC head, at the minimal distance from the LINAC collimators. A graphite block, named back cover wall, stands on a slit which allows to open and close the experimental cavity. Fig.\ref{fig:foto-sorgente} shows a picture of the facility. The LINAC arm is rotated at 90$^{\circ}$ so that the beam runs parallel to the floor. 
The final assembly has a volume of about 1 m$^3$ and weights 1.2 tons. 

\section{The experimental characterization of the e\_LiBANS facility}
\label{sec:measurements}
The neutron field inside the experimental cavity has been accurately characterized by means of multiple measurement techniques. 
The following paragraphs describe the four main measurements performed and the related experimental methods. Section 5 illustrates the results obtained. All the measurements have been taken in the ``closed cavity'' configuration. 

\subsection{The thermal neutron fluence}
\label{sec:Aufoils}
The reference value for the thermal neutron fluence rate in the middle of the irradiation cavity has been firstly measured with standard gold activation foils \cite{Au-foils}. To exclude the contribution from the non-thermal components the cadmium technique has been implemented. Two couples of gold foils have been exposed simultaneously around the central point of the cavity. For each couple, one of the foils was inserted in a small cadmium cover with an average thickness of 0.55 mm and the other was exposed bare. The pure thermal component of the fluence rate is obtained by subtracting the activation due to the non-thermal component (cadmium covered foil) to the total activation of the bare foil. The $^{198}$Au activity, induced by neutron irradiation, has been measured with a High Purity Germanium (HPGe) gamma ray spectroscopy system. The thermal neutron fluence rate, known the activities at saturation $A_{bare}$ and $A_{Cd}$, has been calculated as:
\begin{equation}
\dot\Phi_{W}(th) =\frac{F_c (A_{bare} - \frac{F_b}{F_a} A_{Cd})}{N_{Au} \cdot \sigma_{0} \cdot{g}}
\end{equation}
where $\sigma_0$ is the nominal cross section for thermal neutron capture in gold (equal to 98.69 b at 0.025 eV), $N_{Au}$ the atom density and g is the Westcott factor which takes into account the deviation of the Au cross section from the ideal 1/v dependency.  

The factors F$_a$, F$_b$, F$_c$ are corrective coefficients respectively related to: a) the incomplete thermal neutron attenuation in the Cd cover, b) epithermal neutrons attenuation in the Cd cover, c) the self absorption of the thermal neutrons in Au foil. Since these factors values depends on the energy spectrum of the field and on the foils geometry, they have been calculated with appropriate MCNP6 simulations.

\subsection{The neutron field spatial uniformity}
\label{sec:uniformity}
The fluence distribution in the irradiation volume has been determined by means of a small active thermal neutron rate detector (TNRD) \cite{tnrd} developed within the collaboration. This device, through dedicated acquisition electronics, produces a DC voltage level that is proportional to the thermal neutron fluence rate, and thanks to an internal compensation it is almost insensitive to photons. A calibration of the detector in a known thermal neutron field at a reactor \cite{tnrd-calib} allows to have the real-time value of the measured fluence rate.  
The homogeneity of the thermal neutron field has been measured by moving the detector in 49 different positions following a 7x7 grid on the central cross plane. The fluence rate gradient has been calculated by measuring in 10 positions along the beam axis inside the cavity. 
\subsection{The neutron energy spectrum}
\label{sec:spectrum}
A Bonner Sphere System (BSS) equipped with a TNRD has been used to measure the fluence energy spectrum. Ten high density polyethylene spheres, with diameter ranging from 60 mm to 200 mm, have been exposed in the irradiation volume. The spectrometric system
had been previously calibrated as described in \cite{BSS-tnrd-calib}. To extract the energy spectrum from the detector counts an unfolding procedure has been followed based on the FRUIT unfolding code \cite{fruit,fruit2}. 
The use of the TNRD combined with the BSS allows to operate it in pulsed fields where detectors of other kinds are not effective. 
\subsection{The gamma background}
\label{sec:gamma}
The value of the gamma background in the irradiation volume has been measured with a commercial, energy compensated Geiger Muller (GM) counter insensitive to the neutrons thanks to a proper lithium loaded cap. The 0.64 cm long active volume of the detector has been placed in the middle of the cavity and a bias voltage of 500 kV has been applied. 

\section{Measurement results}
\label{sec:results}

Unless otherwise indicated, all the results described hereinafter have been obtained in the same LINAC working configuration which corresponds to 18 MV photon beam, with 40x40 cm$^2$ collimators aperture and a dose rate equal to 400 Monitor Units\footnote{As for hospital accelerators, the LINAC calibration ensures that 1 MU corresponds to the amount of radiation necessary to deliver 1 cGy of gamma dose at the build up in a standard water phantom at 100 cm from the target.} (MU) per minute.

The value of the thermal neutron fluence rate in the center of the irradiation volume resulting from the Au activation foil measurement is:

$$
\dot{\Phi}_{w}(th) = ( 1.82 \pm 0.04 ) \cdot 10^6 \;\; cm^{-2}s^{-1} 
$$

This value is expressed in terms of \textit{sub-cadmium cut-off fluence rate in the Westcott convention} \cite{Au-foils,Westcott}. 
The associated error represents one standard deviation and is calculated by propagating the uncertainties on the foils irradiation time, the activity measurements and the correction factors.
The difference between the two pairs of foils, simultaneously exposed, is less than 1$\%$.

The cadmium ratio, defined as the activity of the bare foil divided by the activity of the cadmium covered foil, gives an indication of the field energy spectrum. 
For the measured field it results: $$ \frac{A_{bare}}{A_{Cd}} = 6.61 $$ which, as expected, indicates a highly thermalized field.

\begin{figure}
\centering
  \includegraphics[width=0.8\linewidth]{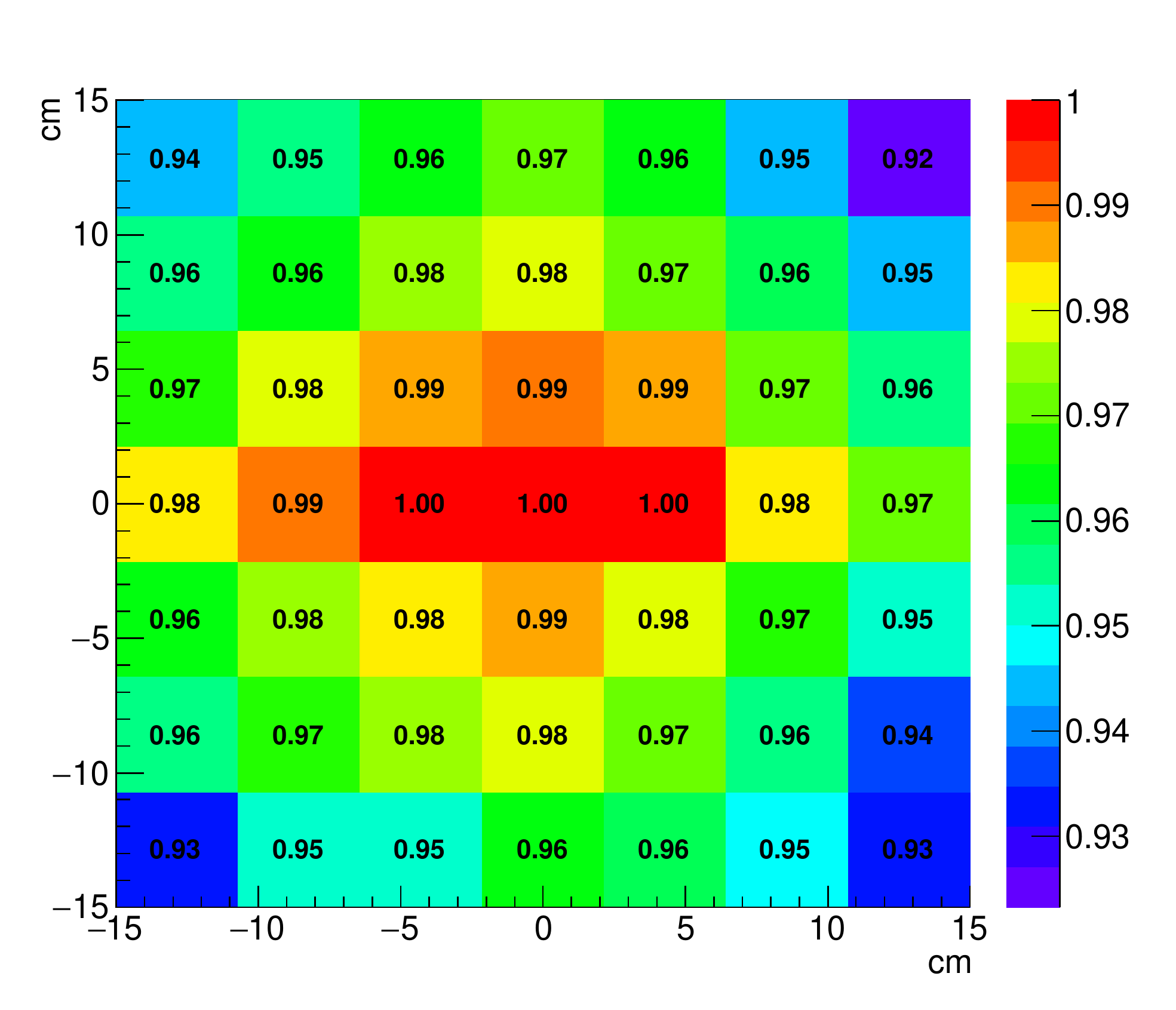}  
 \caption{Uniformity of the measured thermal neutron fluence rate in the cavity transverse plane. Data are normalized to their maximum value, which corresponds to the central position.}
 \label{fig:planarita2}
\end{figure}

The uniformity of the thermal fluence in the transverse section of the irradiation volume is shown in Fig. \ref{fig:planarita2}. Values are normalized to their maximum which corresponds to the central cell.  

A very homogenous field results from the map. Three regions can be analyzed: the complete plane (region A, 49 cells), the complete plane excluding the most external perimeter (region B, 25 cells) and the central region (region C, 9 cells). Tab. \ref{tab:1} records the average value and the standard deviation of the fluence rate distribution in each area together with the maximum discrepancy between the cells in the region.

\begin{table}
\caption{Results of the thermal neutron fluence rate homogeneity measurement in the photoconverter cavity. The first column indicates the average fluence rate in the region, the second column indicates the standard deviation of the measured values divided by the average. The third column shows the maximum discrepancy of the fluence rate values divided by the maximum value.}
\label{tab:1} 
\begin{tabular}{l l l l}
\hline\noalign{\smallskip}
Region & Average fluence (cm$^{-2}$s$^{-1}$) & Standard deviation & Max discrepancy  \\
\noalign{\smallskip}\hline\noalign{\smallskip}
A & $(1.69 \pm 0.03) \cdot 10^6 $ & 1.9\% & 8\% \\
B & $(1.70 \pm 0.03) \cdot 10^6 $ & 1.4\% & 6\% \\
C & $(1.72 \pm 0.03) \cdot 10^6 $ & 0.8\% & 2\% \\
\noalign{\smallskip}\hline
\end{tabular}
\end{table}

The \textit{sub-cadmium fluence rate in the Westcott convention} measured in the middle of the cavity with the TNRD detector, averaged over 4 repeated acquisitions, is:
$$\dot\Phi_{th,W} = (1.74 \pm 0.03) \cdot 10^6 \;\;\; cm^{-2}s^{-1} $$
The error indicates one standard deviation and takes into account the uncertainty on the detector calibration factor, 2\%, \cite{tnrd}, instrumental error, less than 1\%, confirmed by repeated measurements, and the statistical uncertainty, 1.5\%, due to the RMS of the voltage signal and to the baseline subtraction. The accuracy of the result is impressive and it confirms the quality the experimental technique.

\begin{figure}
 \includegraphics[width=0.8\linewidth]{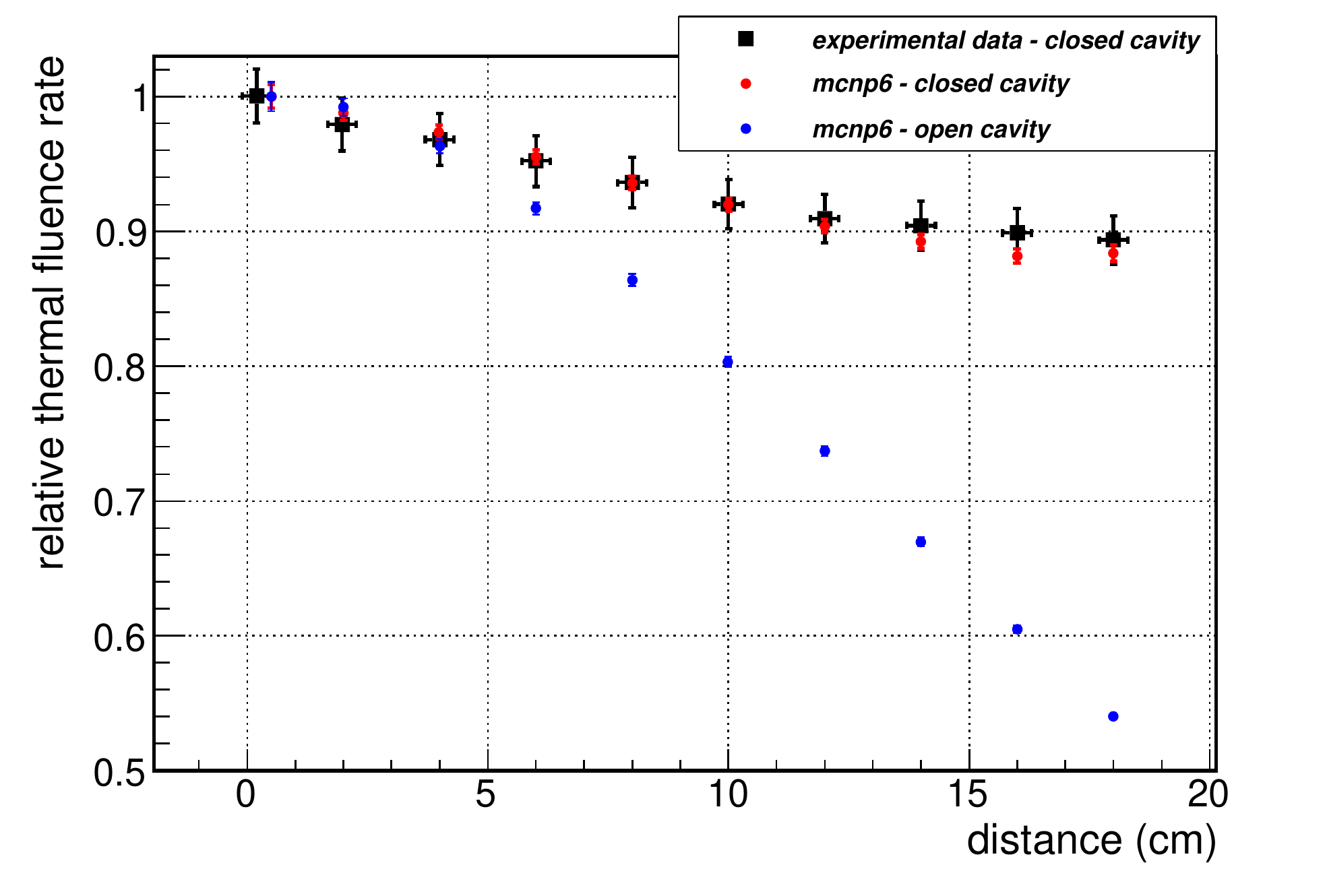} 
 \centering
 \caption{Longitudinal profile of the thermal neutron fluence rate along the central axis of the cavity. X-axis indicates the 
distance from the cavity bottom. Black squared symbols represent the experimental data in the ``closed cavity'' configuration and the circles the simulated values in the same configuration. The triangular symbols correspond to the simulated profile in the ``open cavity'' configuration. }
 \label{fig:gradiente1}
\end{figure}

The longitudinal profile of the thermal neutron fluence rate along the central axis of the cavity is shown in Fig.\ref{fig:gradiente1}. The TNRD has been moved in steps of 2 cm from the bottom of the irradiation volume, indicated with 0 cm, to the most external part, 20 cm. 
As expected the thermal fluence rate goes down getting far from the target in agreement with the MCNP6 simulation.
The reflective effect of the back cover wall is well distinguishable in the flatness of the last points. 
In an ``open cavity'' configuration the rate profile would show much steeper decrease as shown in the plot (label MCNP6 open-cavity).  
The thermal fluence rate decreases by 10.4\% over the full length of the cavity along the central axis, corresponding to an average gradient of 0.56\%/cm. More in detail, it can be observed that the average gradient of the first half of the profile is 0.76\%/cm whilst for the farther points it results 0.38\%/cm. In the error bars, the positioning error of $\pm$0.3 cm is also considered.

\begin{figure}
\centering
  \includegraphics[width=0.8\linewidth]{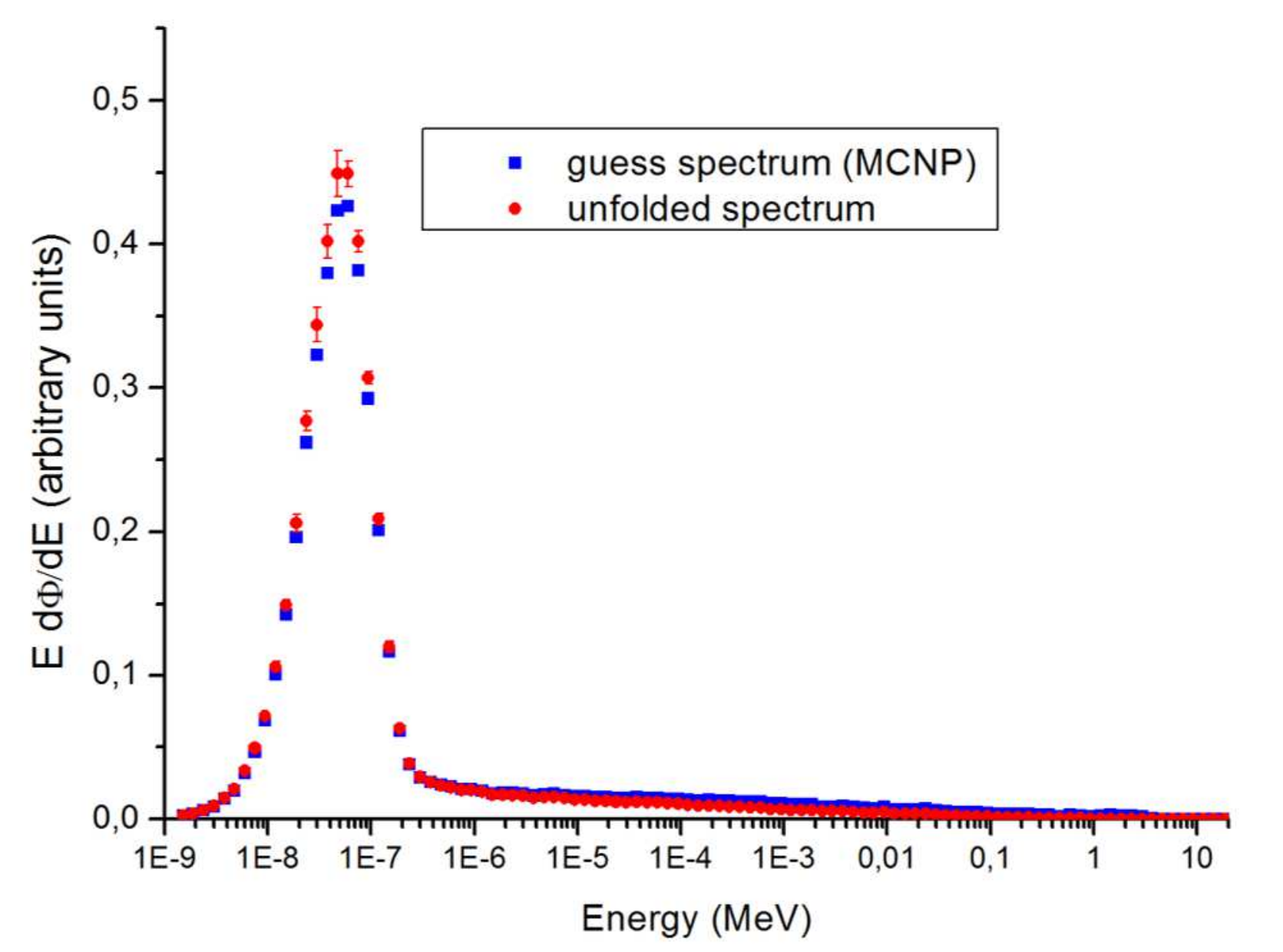}  
 \caption{Neutron energy spectrum measured in the photoconverter cavity. Circles indicate the unfolded data and squared points the MCNP guess. The spectra are normalized to unit fluence.}
 \label{fig:spettro}
\end{figure}

Fig.\ref{fig:spettro} shows the fluence energy spectrum obtained with the Bonner Sphere Spectrometer inside the irradiation volume.
The experimental curve is compared with the MCNP6 calculated curve, which is used as guess spectrum for the numerical unfolding. 
A very well thermalized neutron field is obtained in the cavity, 87\% of the total fluence results below the Cd cut-off and only 2\% populates the fast range above 10 keV.

As far as the gamma background is concerned the gamma dose rate measured with the GM counter positioned in the center of the cavity results: 
$$
D_{\gamma}=1.85\pm 0.08 \; \; \mu Gy/s  
$$
The ratio between the gamma absorbed dose and the thermal neutron fluence rate, often taken as figure of merit for the neutron field purity, is equal to 1.03 10$^{-12}$ Gy cm$^{-2}$, indicating a good shielding against gamma radiation and a pure thermal neutron field. 

\begin{figure}
\centering
  \includegraphics[width=0.8\linewidth]{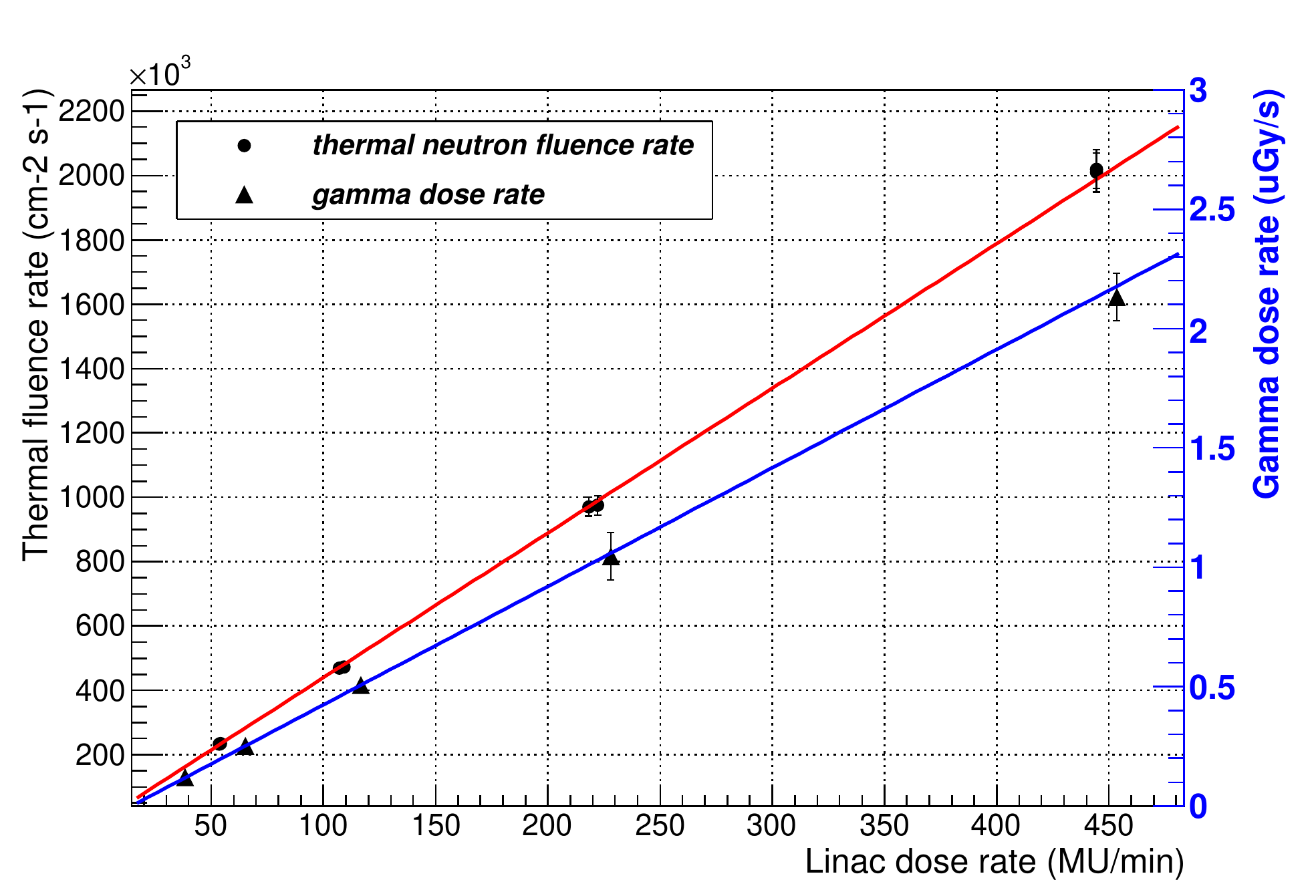}  
 \caption{Linearity of the thermal neutron fluence rate (left) and of the gamma absorbed dose rate (right) in the cavity of the photoconverter as a function of the LINAC dose rate}
 \label{fig:linearita-gamma}
\end{figure}

The neutron and the photon fluence rate are both linearly dependent on the LINAC current. This has been proved through repeated measurements with the TNRD and with the GM counter measuring at different LINAC dose rate, \textit{e.g.} the LINAC current, expressed in terms of MU per minutes. 
The linearity of the thermal neutron fluence rate and of the gamma dose rate are shown in Fig.\ref{fig:linearita-gamma} well fitted by a linear function.

\section{Conclusions}
\label{sec:conclusions}
The new thermal neutron source developed and installed at the Physics Department of University of Torino is based on photonuclear ($\gamma$,n) conversion, obtained through a medical LINAC coupled to a photoconverter.   

The optimal structure for the neutron production and moderation, emerging from MCNP6 simulations and cross section analysis, consists in a lead target and a 
proper graphite and heavy water moderator assembly. The thermal neutron field obtained inside the irradiation volume is very well known with an accuracy of 2\%. The measure of the neutron energy spectrum shows the predominance of the thermal component, attested at 87\% of the total neutron fluence, an epithermal component of 11\% and a residual fast neutron component below 2\%. The purity of the thermal neutron field is also guaranteed by the low gamma dose rate measured in the cavity, producing a figure of merit of the order of 10$^{-12}$ Gy cm$^2$. The thermal neutron fluence rate is homogeneously distributed in the irradiation volume: the maximum deviation is within the 8\% in the cross plane with a slow decrease along the longitudinal axes, 0.6\% cm$^{-1}$ when operated in the ``closed cavity'' configuration.

The e\_LiBANS source offers a small scale irradiation facility with the important feature of the tunability of the thermal fluence rate with the LINAC current intensity. At the present standard working condition the thermal neutron fluence rate is about 2 10$^6$ cm$^{-2}$s$^{-1}$.
The easy access and the reduced costs make this source an attractive solution for various applications needing samples irradiation, like  detectors development and materials study or cell specimens irradiation for preclinical BNCT research where high purity sources are required.

The positive outcome of this work is not only the availability of a new irradiation facility but also the demonstration of the feasibility of this type of source. 
The linear dependence of the output neutron fluence rate on the accelerator current intensity could make this kind of facility competitive with the more complex nuclear reactor extracted beams. 
A dedicated, more powerful accelerator would bring an advantageous increase in the fluence rate intensity that would make the source competitive with nuclear research reactors. Furthermore, the innovative detectors used for this study make this work an important example of complete experimental characterization of a neutron field.

\section*{Acknowledgment}
The project has been supported by Compagnia di San Paolo grant "OPEN ACCES LABS" (2015), Fondazione CRT grant n.2015.AI1430.U1925, INFN CSN 5, MIUR Dipartimenti di Eccellenza (ex L. 232/2016, art. 1, cc. 314, 337).
The authors are also grateful to Azienda Sanitaria Ospedaliera San Luigi Gonzaga - Orbassano, Ospedale S. Giovanni Antica Sede - Torino and to the Elekta S.p.A. for the technical support in the LINAC commissioning and maintenance. 

\section*{References}

\bibliography{vm}

\end{document}